\documentclass[letter,twocolumn]{jpsj3}
\usepackage{txfonts}
\usepackage[dvipdfmx]{graphicx}
\usepackage{bm}
\usepackage{braket}
\usepackage{mathtools}
\usepackage{color}

\title{Type-III Dirac Cones from Degenerate Directionally Flat Bands: 
Viewpoint from Molecular-Orbital Representation}

\author{Tomonari Mizoguchi\thanks{mizoguchi@rhodia.ph.tsukuba.ac.jp}  and Yasuhiro Hatsugai}
\inst{Department of Physics, University of Tsukuba, Tsukuba, Ibaraki 305-8571, Japan}

\abst{
We present a systematic method of creating type-III Dirac cones from degenerate directionally flat bands.
Here, the directionally flat band means a completely dispersionless band only in a certain direction.
The key strategy is to retain one of the directionally flat bands while making the other dispersive, 
by adding a perturbation having a selected form of a Hamiltonian matrix.
Such a form of a perturbation is found by using the 
``molecular-orbital representation" which we have developed to describe flat-band models.
The models thus obtained host type-III Dirac cones even without fine-tuning hopping parameters.
To demonstrate the ubiquity of this method, we study concrete examples of two-band models and a four-band model.}

\begin{document}
\maketitle
\textit{Introduction.--}
Condensed matter realization of Dirac fermions has been of particular interest for decades.
Successful fabrication of graphene~\cite{Wallace1947,Novoselov2004,CastroNeto2009} boosted this research field.
Nowadays, it is found that the Dirac fermions are ubiquitously realized in various materials, both in (quasi-) two dimensions (2D)~\cite{Katayama2006,Goerbig2008,Hirata2016,Ran2009,Morinari2010} and three dimensions (3D)~\cite{Wolff1964,Fuseya2015,Kariyado2011,Liu2014,Borisenko2014,Murakami2007,Burkov2011,Vafek2014,Armitage2018,Bernevig2018}.
 
One of the characteristic features of Dirac fermions in condensed matter systems
is the spatial anisotropy, originating from lattice structures and/or orbital characters. 
Remarkably, the shape of the Fermi surface can be changed while 
keeping the band touching point between conduction and valence bands. 
Specifically, there are three types of Dirac fermions: type-I, type-II, and type-III.
In 2D, the classification of these three is associated with the shape of equi-energy surface.
For the type-I, the equi-energy 
surface around the Dirac point is shaped in ellipse. 
For the type-II, the equi-energy surface around the Dirac point turns to hyperbola. 
The type-III is a critical point between the type-I and type-II,
where one of the bands forming the cone is completely flat in a certain direction,
resulting in the straight-line-shaped Fermi surface.
We note that this classification also holds in 3D, although possible shapes of equi-energy surface may be more abundant. 
From this definition, one finds that the type-III is rare compared with the types I and II, 
but it started to attract attentions recently~\cite{Volovik2016,Volovik2017,Volovik2018,Liu2018,Huang2018,Fragkos2019,Milicevic2019,Kim2020,Chen2020,Jin2020,Gong2020}, in particular, 
as an analog of black-hole event horizon in condensed matters~\cite{Volovik2016,Volovik2017,Volovik2018,Hashimoto2019}. 

Considering the fact that type-I and II Dirac fermions exhibit rich physics, such as
orbital magnetism~\cite{Fukuyama1970,Koshino2007,GomezSantos2011,Raoux2015,Ogata2016}, 
correlations effects (both short-ranged and long-ranged ones)~\cite{Otsuka2002,Son2007,Meng2010,Kotov2012,Isobe2012,Otsuka2016}, 
magnetic-field-induced phenomena~\cite{Abrikosov1998,Novoselov2006,Hatsugai2006,Fukushima2008,Tajima2013,Hatsugai2015_LL},
impurity effects~\cite{Hatsugai1993,Nomura2007,Kanao2012}, 
superconductivity~\cite{Kopnin2008,Mizoguchi2015}, 
non-Hermiticity~\cite{Kozii2017,Yoshida2018,Okugawa2019,Budich2019,Yoshida2019,Papaj2019}, etc., 
type-III Dirac fermions possibly show intriguing physical properties as well.
For example, it was reported that, when considering the superconductivity, 
the superconducting gap is maximized for type-III Dirac fermions
upon changing from type-I to type-II~\cite{Li2017}.
For further studies, search for simple lattice models hosting type-III Dirac cones is highly desirable. 
Such models are also suitable for artificial-materials realization.
However, it is naively expected that one needs quantitative fine-tuning of parameters
to obtain the type-III Dirac fermions, since it is a critical point between the tilted type-I and the type-II.
Then, the following question arises: Can we construct lattice models of the type-III Dirac cones in a systematic manner?

In this letter, we propose a systematic scheme to construct lattice models hosting type-III Dirac fermions.
Our main idea relies on the molecular-orbital (MO) representation~\cite{Hatsugai2011,Hatsugai2015,Mizoguchi2019,Mizoguchi2020}, 
which we have developed to understand and design flat-band models.
Considering the fact that the type-III Dirac cones host the flat branch in a certain direction, 
it is natural that the scheme to describe flat-band models can be utilized to tailor the type-III Dirac cones. 

\begin{figure}[b]
\begin{center}
\includegraphics[clip,width = \linewidth]{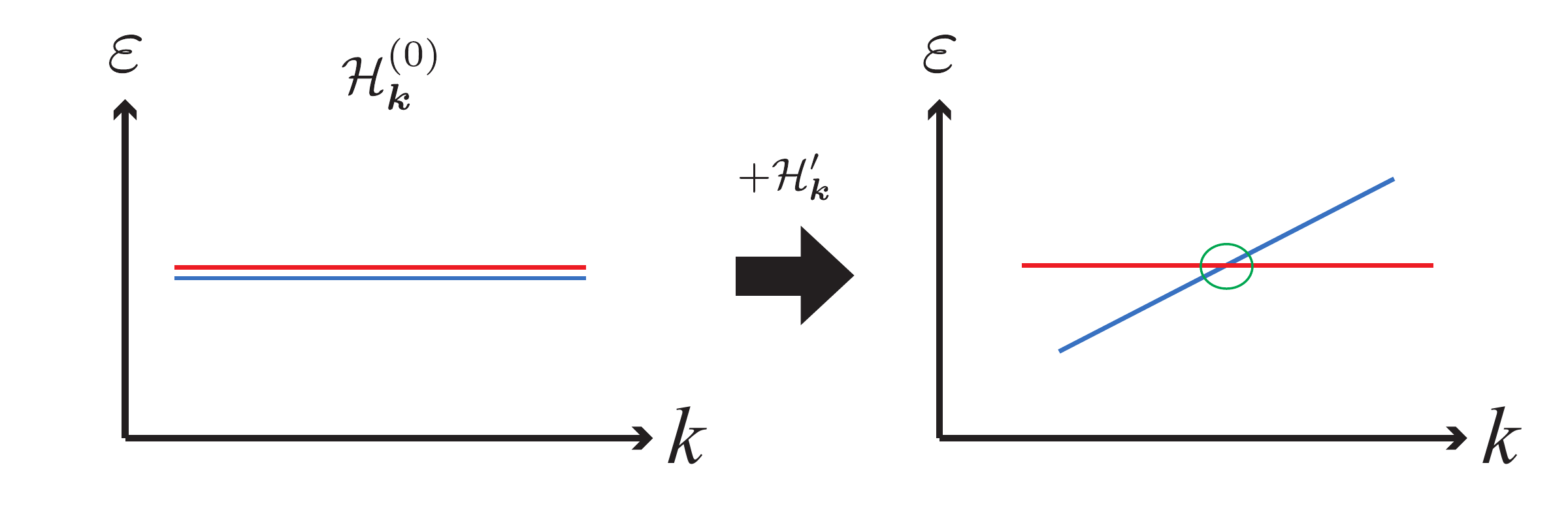}
\vspace{-10pt}
\caption{(Color online) Schematic figure of our method of creating type-III Dirac cones.
The left panel represents the degenerate directionally flat bands and the right panel 
represents the type-III Dirac cone obtained by making the blue branch dispersive while keeping the red branch flat.}
  \label{fig:schematic}
 \end{center}
 \vspace{-10pt}
\end{figure}

The model construction discussed here consists of two steps.
Firstly, we start from the models where zero modes present on certain lines in the Brillouin zone,
which we call a ``directionally flat band", and these modes are doubly degenerate.  
Secondly, we add a perturbation such that one of the degenerate directionally flat bands remains to be flat 
while the other penetrates this band, the crossing point of which is nothing but a type-III Dirac cone (Fig.~\ref{fig:schematic}). 

We demonstrate how this idea works in both minimal two-band models and models with more than three bands.
We present the concrete examples on square and diamond lattices for the two-band case, 
and the two-dimensional Su-Schrieffer-Heeger (2D SSH) model~\cite{Koshino2014,Liu2017,Obana2019,Benalcazar2020} for the four-band case.
In these examples, though the form of the matrix is rather restricted, 
fine-tuning of parameters is not needed to obtain the type-III Dirac fermions.
Therefore, our scheme will open up 
a way to investigation of the physical properties of the type-III Dirac fermions on simple lattice models,
and realization of these models in artificial topological systems such as mechanical systems, electric circuits, photonic crystals and phononic crystals. 
In addition, the viewpoint of the MO representation will be useful to electronic structure analyses on candidate materials.

\textit{Two-band models.---}
Let us first consider the two-band models. 
We begin our argument with a conventional form of Dirac Hamiltonian, 
$\mathcal{H}^{(0)}_{\bm{k}} = \bm{d}_{\bm{k}}
\cdot \bm{\sigma}$,
where $\bm{k}$ denotes the wave vector in 2D or 3D, and 
$\bm{\sigma} = (\sigma_x, \sigma_y, \sigma_z)$ are Pauli matrices. 
We assume the existence of 
the doubly-degenerate zero mode along certain straight lines in the Brillouin zone, labeled by $L_j$.
We refer to this band structure the degenerate directionally flat band.
For the Dirac Hamiltonian, this means that $\bm{d}_{\bm{k}}= 0$ on $\bm{k} \in L_j$.
Although this is a rather restricted situation, 
there exist several well-known examples, as we will show later.
\begin{figure}[b]
\begin{center}
\includegraphics[clip,width = \linewidth]{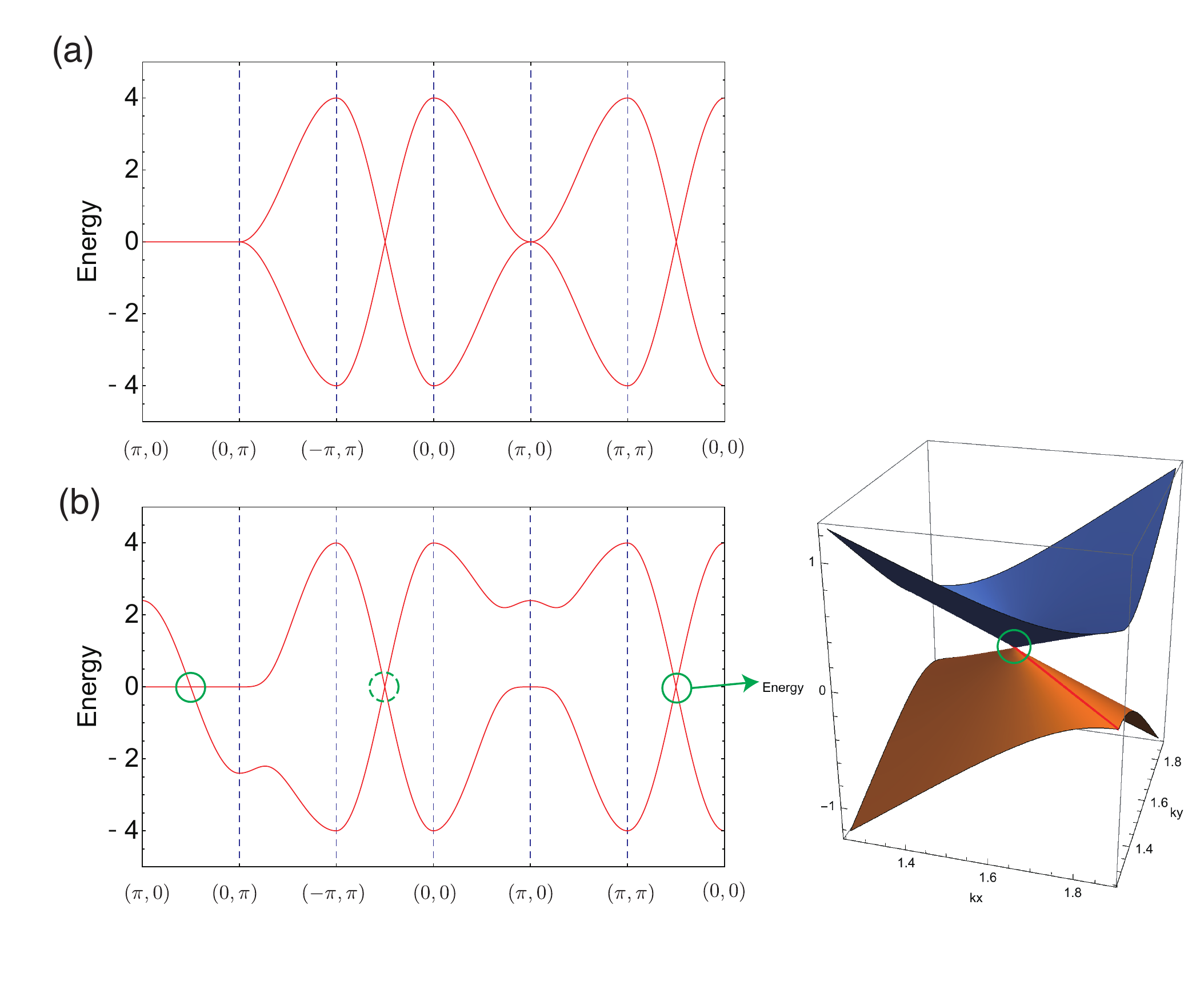}
\vspace{-10pt}
\caption{(Color online) (a) The band structure of the square-lattice model with 
the spin-dependent nearest-neighbor hopping with $t=-1$. 
(b) The band structure for the model with the additional nearest-neighbor hoppings with 
$t=-1$ and $t^\prime = -0.3$.
The green solid and dashed circles denote the type-III Dirac cones, 
on the lines of $k_x + k_y = \pi$ and $k_x + k_y = - \pi$, respectively,
and the red line in the 3D plot is the flat branch of the Dirac cone. }
  \label{fig:square}
 \end{center}
 \vspace{-10pt}
\end{figure}

To create the type-III Dirac cones, we need to modify the model
such that one of the degenerate directionally flat bands is remained to be flat, 
while the other is dispersive and intersects the directionally flat band.
To do this, we modify the Hamiltonian as
$\mathcal{H}^{\rm tot}_{\bm{k}} =  \mathcal{H}^{(0)}_{\bm{k}} +\mathcal{H}^\prime_{\bm{k}}$,
where
\begin{eqnarray}
\mathcal{H}^\prime_{\bm{k}} = 
\begin{pmatrix}
a_{\bm{k}}  & a_{\bm{k}} e^{-i\phi_{\bm{k}}} \\ 
a_{\bm{k}} e^{i\phi_{\bm{k}}}  & a_{\bm{k}} \\ 
\end{pmatrix},
\end{eqnarray}
with $a_{\bm{k}} $ and $\phi_{\bm{k}}$ being the real functions of $\bm{k}$.
In fact, $\mathcal{H}^\prime_{\bm{k}}$ is a ``flat-band" Hamiltonian.
To be specific, $\mathcal{H}^\prime_{\bm{k}}$ can be written 
by using a single vector:
\begin{eqnarray}
\mathcal{H}^\prime_{\bm{k}} = 
a_{\bm{k}} \bm{x}_0 \bm{x}_0^{\dagger}, \label{eq:Hp_MOrep}
\end{eqnarray}
with 
$\bm{x}_0 = \left(1, e^{i\phi_{\bm{k}}} \right)^{\rm T}$.
Equation~(\ref{eq:Hp_MOrep}) implies that the rank of the Hamiltonian is one rather than two,
because the vector $(1,-e^{i\phi_{\bm{k}}})^{\rm T}$, which is orthogonal to $\bm{x}_0^\dagger$, 
becomes the zero-energy eigenvector.
More generally, when the $N \times N$ Hamiltonian matrix is 
written as $\mathcal{H}_{\bm{k}} = \Psi_{\bm{k}} h_{\bm{k}} \Psi_{\bm{k}}^\dagger$
where $\Psi_{\bm{k}}$ and $h_{\bm{k}}$ are $N \times M $ and $M \times M $ matrices, respectively (with $M<N$),
the existence of $N-M$ zero modes is guaranteed regardless of $\bm{k}$ 
because the dimension of the kernel of $\Psi_{\bm{k}}^\dagger$ is greater than or equal to $N-M$.
Such a representation of the Hamiltonian matrix by using the non-square matrix
is called the MO representation, and we call each column of $\Psi_k$ a MO since 
it typically corresponds to a linear combination of atomic orbitals. 
In fact, the MO representation is useful to study and design flat-band models~\cite{Hatsugai2011,Hatsugai2015,Mizoguchi2019,Mizoguchi2020}. 
In the following, for simplicity, we consider the case of $\phi_{\bm{k}} =0$.

On $\bm{k} \in L_j$, 
we have $\mathcal{H}^{\rm tot}_{\bm{k}} = \mathcal{H}^\prime_{\bm{k}}$ as $\mathcal{H}^{(0)}_{\bm{k}} = 0$.
Then, the dispersion relation along this line becomes
$\varepsilon_{\bm{k}} = 0, 2a_{\bm{k}}$. 
This means that, if the sign change of $a_{\bm{k}}$ occurs at a certain point on $L_j$, 
that point naturally becomes a type-III Dirac point.
Notably, $a_{\bm{k}}$ can be chosen flexibly as far as this condition is satisfied, 
so the fine-tuning of parameters is not needed. 
\begin{figure}[t]
\begin{center}
\includegraphics[clip,width = \linewidth]{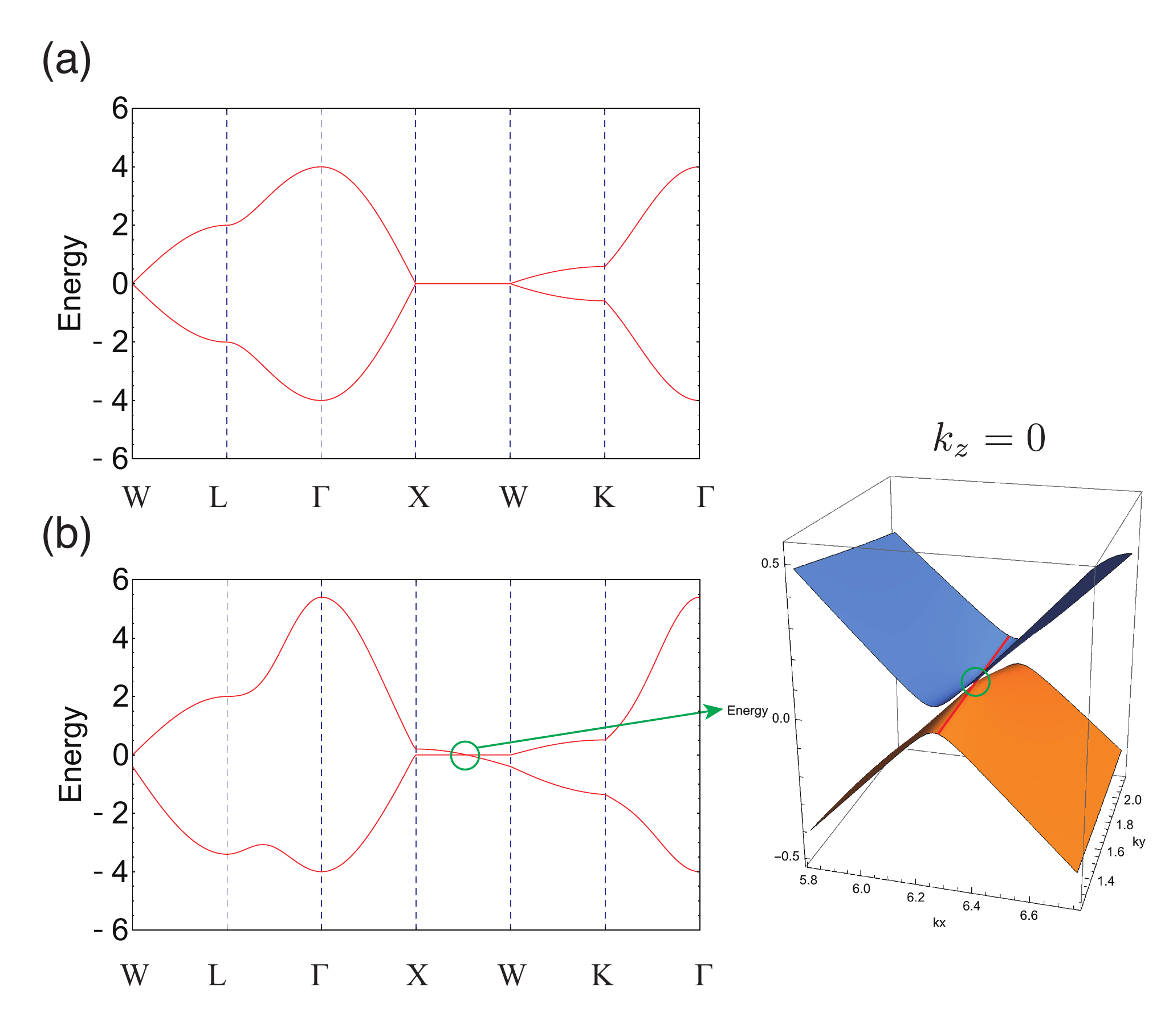}
\vspace{-10pt}
\caption{(Color online) 
(a) The band structure of the nearest-neighbor hopping model on a diamond lattice with $t=1$.
(b) The band structure of the modified model with 
$(t,\lambda_1,\lambda_2,\lambda_3) = (1,0.2,0.1,0.05)$.
The green circle denotes the type-III Weyl point, and the red line in the 3D plot is the flat branch.
The coordinates of the high-symmetry points are 
$\Gamma = (0,0,0)$, 
$\mathrm{X} =\left(2\pi,0,0 \right)$,
$ \mathrm{W}= \left(2\pi, \pi, 0 \right)$,
$ \mathrm{K}= \left( \frac{3\pi}{2},\frac{3\pi}{2},0\right)$,
and 
$ \mathrm{L}= \left(\pi, \pi, \pi \right)$.}
  \label{fig:diamond}
 \end{center}
 \vspace{-10pt}
\end{figure}

We demonstrate how this idea works in the concrete examples.
The first example is the square-lattice model in 2D for spinful fermions.
The original Hamiltonian $\mathcal{H}^{(0)}_{\bm{k}}$ is chosen to be the nearest-neighbor hopping model,
but the sign of the transfer depends on the spin~\cite{Kudo2019}.
Namely, we take $\bm{d}_{\bm{k}} = (0,0,d^z_{\bm{k}})$ with $d^z_{\bm{k}} = 2t \left( cos k_x + \cos k_y \right)$.
$d^z_{\bm{k}}$ becomes zero for the lines $k_x + k_y = \pm \pi$, i.e., the directionally flat band is formed on these lines [Fig.~\ref{fig:square}(a)].

As for the additional term, we set  $a_{\bm{k}}$ as 
\begin{eqnarray}
a_{\bm{k}} = 2t^\prime \left( cos k_x - \cos k_y  \right).
\end{eqnarray}
It is, however, to be emphasized that $a_{\bm{k}}$ is not necessarily chosen in this form.
We also note that $\mathcal{H}^\prime_{\bm{k}}$ 
corresponds to the nearest-neighbor hoppings whose sign is opposite in $x$ direction is opposite to 
that in $y$ direction,
for both spin-conserved and spin-flipping channels. 
$a_{\bm{k}}$ becomes zero when $k_x  = \pm k_y$.
Then, from our generic argument, the type-III Dirac cones appear at the intersections of 
$k_x + k_y = \pm \pi$ and $k_x  = \pm k_y$, that are, 
$\bm{k} = \left( \pm \frac{\pi}{2}, \pm \frac{\pi}{2}\right)$.
In Fig.~\ref{fig:square}(b), we plot the band structure for the model 
$\mathcal{H}^{(0)}_{\bm{k}} + \mathcal{H}^\prime_{\bm{k}}$.
We indeed find the type-III Dirac cones at those points.

The second example is the diamond-lattice model in 3D for spinless fermions,
where we can create the type-III Weyl semimetal. 
The original Hamiltonian $\mathcal{H}^{(0)}_{\bm{k}}$ is again chosen to be the simple nearest-neighbor hopping model.
The corresponding $\bm{d}_{\bm{k}}$ is $(d^x_{\bm{k}}, d^y_{\bm{k}},0)$ with 
$d^x_{\bm{k}} = t \left(1 + \cos \bm{k} \cdot \bm{a}_1  + \cos \bm{k} \cdot \bm{a}_2 + \cos \bm{k} \cdot \bm{a}_3 \right)$ and 
$d^y_{\bm{k}} = t \left(\sin \bm{k} \cdot \bm{a}_1  + \sin \bm{k} \cdot \bm{a}_2 + \sin \bm{k} \cdot \bm{a}_3 \right)$;
here $\bm{a}_1 = \left(0, \frac{1}{2}, \frac{1}{2}\right)$, $\bm{a}_2 = \left( \frac{1}{2}, 0, \frac{1}{2}\right)$,
and $\bm{a}_3 = \left(\frac{1}{2}, \frac{1}{2},0\right)$ are three lattice vectors. 
This model hosts doubly-degenerate directionally flat band on X-W lines, i.e, the line on $(2\pi, k_y,0)$, as shown in Fig.~\ref{fig:diamond}(a). 

To realize the type-III Weyl semimetal, we again add $\mathcal{H}^\prime_{\bm{k}}$. 
In the present model, we choose 
\begin{eqnarray}
a_{\bm{k}}  = 2 \lambda_1 \cos \left( \bm{k} \cdot \bm{a}_1\right) + 2 \lambda_2  \left(\cos \bm{k} \cdot \bm{a}_2\right)  + 2 \lambda_3 \cos \left( \bm{k} \cdot \bm{a}_3\right).
\end{eqnarray}
Although this looks simple, the actual hoppings corresponding to $a_{\bm{k}}$ are more or less unnatural,
namely, they are direction-selected second-neighbor hoppings between the sites of the same sublattice, and 
the nearest-neighbor and the third neighbor hoppings between the sites of the different sublattice.
The band structure of the modified model is in shown in Fig.~\ref{fig:diamond}(b).
We see the Weyl point between X and W, where $a_{\bm{k}}  = 0$ is satisfied. 
Hence, the type-III Weyl semimetal is realized in this model.  

\begin{figure*}[t]
\begin{center}
\includegraphics[clip,width = \linewidth]{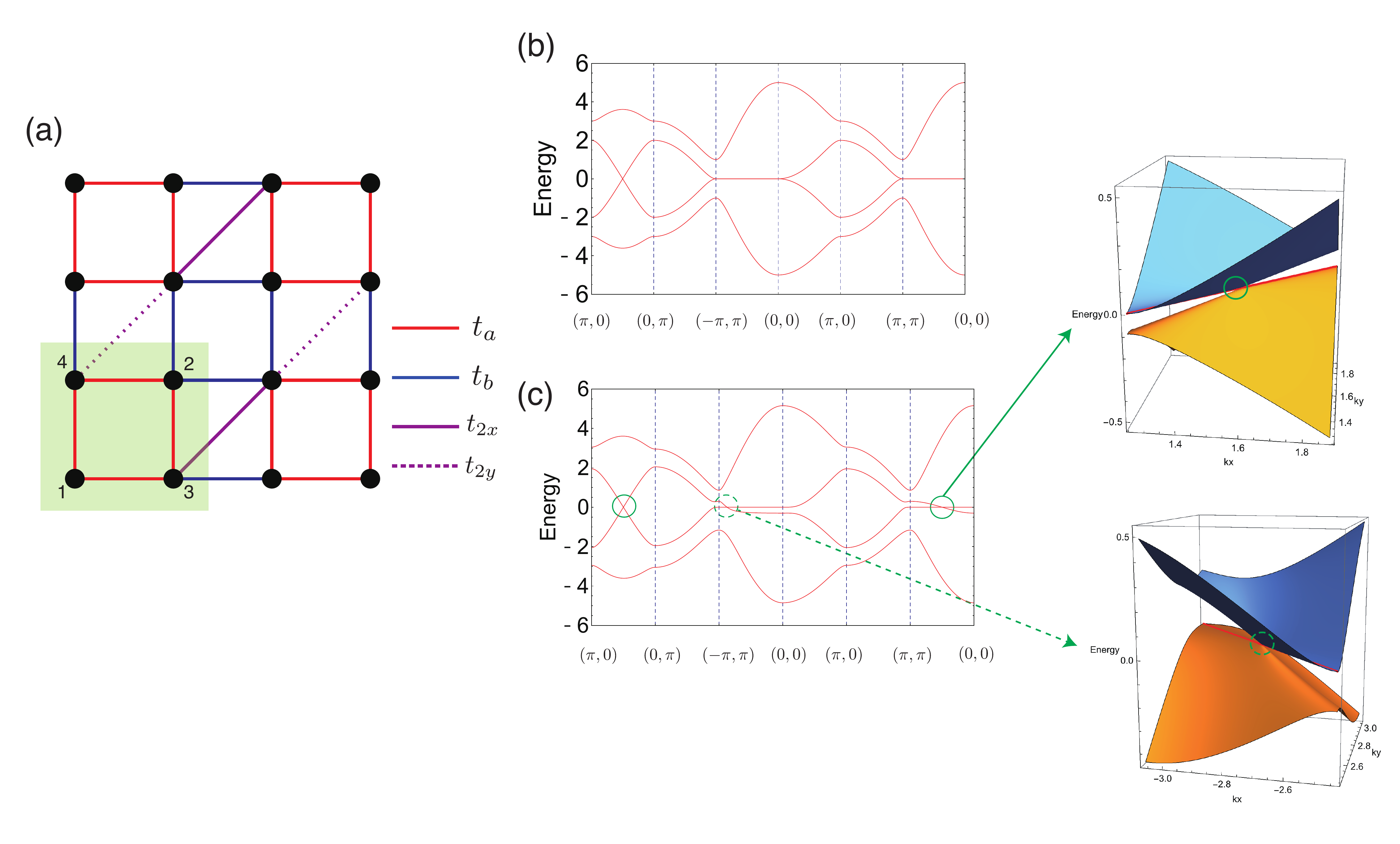}
\vspace{-10pt}
\caption{(Color online) 
(a) Schematic figure of the modified 2D SSH model.
The green shade represents the unit cell.
The hoppings on red and blue bonds are $t_a$ and $t_b$, respectively; 
those on solid and dashed purple bonds are $t_{2x}$ and $t_{2y}$, respectively. 
For the conventional 2D SSH model, $t_{2x} = t_{2y} = 0$. 
(b) The band structure of the 2D SSH model with $t_a=1$, $t_b  =1.5$, $t_{2x} = t_{2y} = 0$
(i.e., the conventional 2D SSH model), and
(c) that of the modified model with $t_a=1$, $t_b  =1.5$, $t_{2x} = 0.2$, and $t_{2y} = 0.1$.
The green solid and dashed circles denote the type-III Dirac cones on the lines of $k_x = k_y$ and $k_x = -k_y$,
respectively,
and the red lines in the 3D plots are the flat branch of the Dirac cone.}
  \label{fig:2dssh_mod}
 \end{center}
 \vspace{-10pt}
\end{figure*}
\textit{Beyond two-band models.---}
Next, we consider the models having more than three internal degrees of freedom,
where the existence of degenerate directionally flat bands 
is not necessarily the consequence of the Hamiltonian matrix being zero.
In other words, the original Hamiltonian itself can be written by the MOs along certain lines.
As a result, the form of the additional term becomes less restricted, which allows us to
construct more feasible models compared with the two-band models. 

For concreteness, let us consider the 2D SSH model~\cite{Koshino2014,Liu2017,Obana2019,Benalcazar2020} [Fig.~\ref{fig:2dssh_mod}(a)], 
whose Hamiltonian matrix is given as
\begin{eqnarray}
\mathcal{H}^{(0)}_{\bm{k}} = 
\begin{pmatrix}
0  & 0 & \alpha_{k_x} & \beta_{k_y} \\
0 & 0 & \beta_{k_y}^\ast & \alpha_{k_x}^\ast \\
\alpha_{k_x}^\ast & \beta_{k_y} & 0 & 0 \\
\beta_{k_x}^\ast & \alpha_{k_y} & 0 & 0 \\
\end{pmatrix}, \label{eq:ham_2dssh}
\end{eqnarray}
where $\alpha_{k_x} = t_a +t_be^{-ik_x}$ and $\beta_{k_y} = t_a +t_be^{-ik_y}$.
In this model, the dispersion relation can be obtained as 
$\varepsilon_{\bm{k}} = \pm \left| |\alpha_{k_x}| \pm |\beta_{k_y}| \right|$.
From this expression, we find that 
there exists a doubly-degenerate zero-energy directionally flat band
in the Brillouin zone where $|\alpha| = |\beta|$ is satisfied~\cite{Liu2017,Obana2019,Benalcazar2020,Koshino2014}.
Specifically, this is satisfied on the lines $k_x = k_y$ and $k_x = -k_y$.
In Fig.~\ref{fig:2dssh_mod}(b), we plot the band structure, 
where we indeed find the doubly-degenerate zero-energy directionally flat band on these lines. 

To proceed, we employ the MO representation along these lines, namely, 
we rewrite the Hamiltonian matrix by using the non-square matrix.
In the following, we focus on the line of $k_x = k_y$, but the similar argument holds for $k_x = -k_y$.
Setting $k_x = k_y = k$, we have $\alpha_{k} = \beta_{k}$,
which we denote $\gamma_k = |\gamma_k| e^{i \theta_k}$.
Note that $|\gamma_k| \neq 0$ for any $k$ if $|t_a| \neq |t_b| $.
Let $\Psi_k$ be a $4 \times 2$ matrix:
\begin{eqnarray}
\Psi_k = \begin{pmatrix}
e^{ i \theta_k} & 0 \\
e^{-i \theta_k}& 0 \\
0 & 1 \\
0 & 1 \\
\end{pmatrix},
\end{eqnarray}
and $h_k$ be a $2 \times 2$ matrix: 
$h_k=  |\gamma_k| \sigma_x$.
Then, the Hamiltonian of Eq.~(\ref{eq:ham_2dssh}) can be rewritten as
\begin{eqnarray}
\mathcal{H}^{(0)}_{(k,k)} = & 
\begin{pmatrix}
0  & 0 & \gamma_k & \gamma_k \\
0 & 0 &\gamma_k^\ast & \gamma_k^\ast \\
\gamma_k^\ast & \gamma_k& 0 & 0 \\
\gamma_k^\ast & \gamma_k& 0 & 0 \\
\end{pmatrix} \nonumber \\
= & \Psi_k h_k \Psi^\dagger_k.
\end{eqnarray}
As the dimension of the kernel of $\Psi_k^\dagger$ is two,
there exist two zero modes of $\mathcal{H}^{(0)}_{(k,k)}$ for any $k$.

Having this MO representation at hand, we now consider 
how to deform the doubly-degenerate flat band into the type-III Dirac cone. 
Our strategy is to add $\mathcal{H}^{\prime}_{\bm{k}}$ such that the Hamiltonian is written in the form
\begin{eqnarray}
\mathcal{H}^{\rm tot}_{(k,k)}  = \mathcal{H}^{(0)}_{(k,k)} + \mathcal{H}^{\prime}_{(k,k)}  
=  \tilde{\Psi}_k \tilde{h}_k \tilde{\Psi}^\dagger_k, \label{eq:MO_modified}
\end{eqnarray}
where 
\begin{eqnarray}
\tilde{\Psi}_k = \begin{pmatrix}
e^{ i \theta_k} & 0 & 0 \\
e^{-i \theta_k}& 0 & 0 \\
0 & 1 &1 \\
0 & 1&-1 \\
\end{pmatrix},
\end{eqnarray}
and 
$\tilde{h}_{k}$ satisfies det $\tilde{h}_{k} = 0$ at some points on $k \in [-\pi,\pi]$;
the second condition ensures that the band crossing between the directionally flat band and a dispersive band occurs~\cite{Mizoguchi2019}.
In $\tilde{\Psi}_k$, a new molecular orbital, $(0,0,1,-1)$, is added to $\Psi_k$.
As $\tilde{\Psi}^\dagger_k$ is a $3 \times 4$ matrix 
and thus dim $[\mathrm{Ker} (\tilde{\Psi}^{\dagger}_k)] = 1$, 
one of the bands remains to be flat and to have zero energy. 
Then, the point satisfying det $\tilde{h}_{k} = 0$ naturally becomes a type-III Dirac cone. 

The above strategy can be realized by adding, for example, the following term:
\begin{eqnarray}
 \mathcal{H}^{\prime}_{\bm{k}}  
= 
\begin{pmatrix}
0 & 0 & 0 &0 \\
0& 0 & 0 & 0 \\
0 & 0 & 0 & t_{2x} e^{ik_x} + t_{2y} e^{-ik_y} \\
0 & 0 &  t_{2x} e^{-ik_x} + t_{2y} e^{ik_y}&0 \\
\end{pmatrix},
\end{eqnarray}
which is schematically depicted by solid and dashed purple lines in
Fig.~\ref{fig:2dssh_mod}(a). 
We again emphasize that this is not the only form of $\mathcal{H}^{\prime}_{\bm{k}}$ that can realize the type-III Dirac cone. 
This additional term corresponds to the inter-unit-cell second-neighbor 
hoppings between the sites of sublattices 3 and 4.
Then, the total Hamiltonian $\mathcal{H}^{\mathrm{tot}}_{(k,k)}$ is indeed 
written in the form of Eq.~(\ref{eq:MO_modified}), and 
the corresponding $\tilde{h}_{k}$ is given as 
\begin{eqnarray}
\tilde{h}_k = 
\begin{pmatrix}
0 & |\gamma_k| &  0 \\
 |\gamma_k| &f_k& i g_k\\
0&-i g_k & - f_k \\
\end{pmatrix}, \label{eq:tildeh}
\end{eqnarray} 
where $f_k = \frac{\left(t_{2x} + t_{2y} \right) \cos k}{2}$ and $g_k = \frac{\left(t_{2x} - t_{2y} \right) \sin k}{2}$. 
From Eq.~(\ref{eq:tildeh}), one finds that det $\tilde{h}_k = 0$ is satisfied when $f_k = 0$~\cite{remark2}, i.e., $t_{2x} + t_{2y} = 0$ or $k =\pm \frac{\pi}{2}$.
For the former case, the doubly degenerate flat band remains along the entire line, 
while the latter case yields the type-III Dirac cone. 

The resulting band structure is shown in Fig.~\ref{fig:2dssh_mod}(c).
As expected, we obtain the type-III Dirac cones on the line of $k_x = k_y$, and 
they also appear on the line of $k_x = -k_y$.
Note that the Dirac cone on the line $k_x = -k_y$ [a green dashed circle in Fig.~\ref{fig:2dssh_mod}(c)] is not located at $k_x = -k_y = \pm \frac{\pi}{2}$.
This is due to the fact that $\mathcal{H}_{\bm{k}}^\prime$ breaks the four-fold rotational symmetry.
From the viewpoint of the MO representation, this is due to the fact that 
the form of $\tilde{h}$ on the line of $k_x = -k_y$ is different from that on $k_x = k_y$.

\textit{Summary.---}
In summary, we have proposed a systematic construction of the type-III Dirac cones for simple lattice models. 
In short, our proposal is summarized as follows: 
it is not the model parameters but the form of the Hamiltonian matrix 
that is to be tuned to create the type-III Dirac cones.
To be concrete, we start from the models having degenerate directionally flat band,
and add a perturbation that can retain one of the directionally flat band while making the other intersect the flat band. 
We demonstrate how this idea works by presenting the examples 
of two-band models, i.e., the square-lattice model with spin-dependent hoppings and the diamond-lattice model,
as well as the 2D SSH model which has four bands.
For two-band cases, we start from the models where the Hamiltonian 
becomes zero along certain directions,
and the additional term to lift the degeneracy of the directionally flat band is represented by the MO.
For the 2D SSH model, the original Hamiltonian itself is represented by the MOs,
thus the type-III Dirac cone is obtained by controlling the number of MOs we need to express the Hamiltonian.  
We hope that our scheme of the model construction is useful to further studies of the type-III Dirac cones
in various aspects, such as electronic structure analyses on candidate materials (e.g., Zn$_2$In$_2$S$_5$), 
numerical studies on simple lattice models and the design of artificial materials.

\begin{acknowledgement}
This work is supported by the JSPS KAKENHI, 
Grant Nos.~JP17H06138 and JP20K14371 (TM), Japan.
\end{acknowledgement}


\begin{thebibliography}{9}
\bibitem{Wallace1947}
P. R. Wallace, Phys. Rev. \textbf{71}, 622 (1947).

\bibitem{Novoselov2004}
K. S. Novoselov, A. K. Geim, S. V. Morozov, D. Jiang, Y. Zhang, S. V. Dubonos, I. V. Grigorieva, and A. A. Firsov, 
Science \textbf{306}, 666 (2004).

\bibitem{CastroNeto2009}
A. H. Castro Neto, F. Guinea, N. M. R. Peres, K. S. Novoselov, and A. K. Geim,
Rev. Mod. Phys. \textbf{81}, 109 (2009). 

\bibitem{Katayama2006}
S. Katayama, A. Kobayashi, and Y. Suzumura, 
J. Phys. Soc. Jpn. \textbf{75}, 054705 (2006). 

\bibitem{Goerbig2008}
M. O. Goerbig, J.-N. Fuchs, G. Montambaux, and F. Pi\'{e}chon,
Phys. Rev. B \textbf{78}, 045415 (2008).

\bibitem{Hirata2016}
M. Hirata, K. Ishikawa, K.Miyagawa, M. Tamura, C. Berthier, D. Basko,
A. Kobayashi, G. Matsuno, and K. Kanoda,
Nat. Commun. \textbf{7}, 12666 (2016).

\bibitem{Ran2009} 
Y. Ran, F. Wang, H. Zhai, A. Vishwanath, and D. H. Lee, 
Phys. Rev. B \textbf{79}, 014505 (2009).

\bibitem{Morinari2010} 
T. Morinari, E. Kaneshita, and T. Tohyama, 
Phys. Rev. Lett. \textbf{105}, 037203 (2010).

\bibitem{Wolff1964}
P. Wolff, 
J. Phys. Chem. Solids \textbf{25}, 1057 (1964).

\bibitem{Fuseya2015}
Y. Fuseya, M. Ogata, and H. Fukuyama,
J. Phys. Soc. Jpn. \textbf{84}, 012001 (2015).

\bibitem{Kariyado2011}
T. Kariyado and M. Ogata, 
J. Phys. Soc. Jpn. \textbf{80}, 083704 (2011); 
J. Phys. Soc. Jpn. \textbf{81}, 064701 (2012).

\bibitem{Liu2014}
Z. K. Liu, B. Zhou, Y. Zhang, Z. J. Wang, H. M. Weng, D. Prabhakaran, S.-K. Mo, Z. X. Shen, Z. Fang, X. Dai, Z. Hussain, and Y. L. Chen, 
Science \textbf{343}, 864 (2014).

\bibitem{Borisenko2014}
S. Borisenko, Q. Gibson, D. Evtushinsky, V. Zabolotnyy, B. B\"{u}chner, and R. J. Cava,
Phys. Rev. Lett. \textbf{113}, 027603 (2014). 

\bibitem{Murakami2007}
S. Murakami, 
New J. Phys. \textbf{9}, 356 (2007).

\bibitem{Burkov2011}
A. A. Burkov and L. Balents,
Phys. Rev. Lett. \textbf{107}, 127205 (2011).

\bibitem{Vafek2014} O. Vafek and A. Vishwanath, 
Annu. Rev. Condens. Matter Phys. \textbf{5}, 83 (2014).  

\bibitem{Armitage2018} N. P. Armitage, E. J. Mele, and A. Vishwanath,
Rev. Mod. Phys. \textbf{90}, 015001 (2018). 

\bibitem{Bernevig2018}
A. Bernevig, H. Weng, Z. Fang, and X. Dai,
J. Phys. Soc. Jpn. \textbf{87}, 041001 (2018).

\bibitem{Volovik2016}
G. E. Volovik, JETP Lett. \textbf{104}, 645 (2016).

\bibitem{Volovik2017}
G. E. Volovik, and K. Zhang,
J. Low Temp. Phys. \textbf{189}, 276 (2017).

\bibitem{Volovik2018}
G. E. Volovik, Phys.-Usp. \textbf{61}, 89 (2018).

\bibitem{Liu2018}
H. Liu, J.-T. Sun, C. Cheng, F. Liu, and S. Meng,
Phys. Rev. Lett. \textbf{120}, 237403 (2018). 

\bibitem{Huang2018}
H. Huang, K.-H. Jin, and F. Liu,
Phys. Rev. B \textbf{98}, 121110(R) (2018).

\bibitem{Fragkos2019}
S. Fragkos, R. Sant, C. Alvarez, E. Golias, J. Marquez-Velasco, P. Tsipas, D. Tsoutsou, S. Aminalragia-Giamini, E.  Xenogiannopoulou, H. Okuno, G. Renaud, O. Rader, and A. Dimoulas,
Phys. Rev. Mater. \textbf{3}, 104201 (2019).

\bibitem{Milicevic2019}
M. Mili\'{c}evi\'{c}, G. Montambaux, T. Ozawa, O. Jamadi, B. Real, I. Sagnes, A. Lema\^{i}tre, L. Le Gratiet, A. Harouri, J. Bloch, and A. Amo,
Phys. Rev. X \textbf{9}, 031010 (2019).

\bibitem{Kim2020}
J. Kim, S. Yu, and N. Park,
Phys. Rev. Applied, \textbf{13}, 044015 (2020).

\bibitem{Chen2020}
Y.-G. Chen, X. Luo, F.-Y. Li, B. Chen, and Y. Yu,
Phys. Rev. B \textbf{101}, 035130 (2020).

\bibitem{Jin2020}
L. Jin, H. C. Wu, B.-B. Wei, and Z. Song,
Phys. Rev. B \textbf{101}, 045130 (2020).

\bibitem{Gong2020}
Z. Gong, X. Shi, J. Li, S. Li, C. He, T. Ouyang, 
C. Zhang, C. Tang, and J. Zhong,
Phys. Rev. B \textbf{101}, 155427 (2020).

\bibitem{Hashimoto2019}
K. Hashimoto and Y. Matsuo,
arXiv:1911.04675.

\bibitem{Fukuyama1970}
H. Fukuyama and R. Kubo,
J. Phys. Soc. Jpn. \textbf{28}, 570 (1970).

\bibitem{Koshino2007}
M. Koshino and T. Ando, 
Phys. Rev. B \textbf{76}, 085425 (2007);
Phys. Rev. B \textbf{81}, 195431 (2010).

\bibitem{GomezSantos2011}
G. G\'omez-Santos and T. Stauber,
Phys. Rev. Lett. \textbf{106}, 045504 (2011).

\bibitem{Raoux2015}
A. Raoux, R. Pi\'{e}chon, J.-N. Fuchs, and G. Montambaux,
Phys. Rev. B \textbf{91}, 085120 (2015).

\bibitem{Ogata2016}
M. Ogata, J. Phys. Soc. Jpn. 
\textbf{85}, 104708 (2016).

\bibitem{Otsuka2002}
Y. Otsuka and Y. Hatsugai,
Phys. Rev. B \textbf{65}, 073101 (2002).

\bibitem{Son2007}
D. T. Son, Phys. Rev. B \textbf{75}, 235423 (2007).

\bibitem{Meng2010}
Z. Y. Meng, T. C. Land, S. Wessel, F. F. Assaad, and A. Muramatsu,
Nature (London) \textbf{464}, 847 (2010).

\bibitem{Kotov2012}
V. N. Kotov, B. Uchoa, V. M. Pereira, F. Guinea, and A. H. Castro Neto,
Rev. Mod. Phys. \textbf{84}, 1067 (2012).

\bibitem{Isobe2012}
H. Isobe and N. Nagaosa,
J. Phys. Soc. Jpn. \textbf{81}, 113704 (2012).

\bibitem{Otsuka2016}
Y. Otsuka, S. Yunoki, and S. Sorella,
Phys. Rev. X \textbf{6}, 011029 (2016).

\bibitem{Abrikosov1998}
A. A. Abrikosov, Phys. Rev. B \textbf{58}, 2788 (1998).

\bibitem{Novoselov2006}
K. S. Novoselov, E. McCann, S. V. Morozov, V. I. Fal'ko,
M. I. Katsnelson, U. Zeitler, D. Jiang, F. Schedin, and A. K. Geim,
Nat. Phys. \textbf{2}, 177 (2006).

\bibitem{Hatsugai2006}
Y. Hatsugai, T. Fukui, and H. Aoki,
Phys. Rev. B \textbf{74}, 205414 (2006). 

\bibitem{Fukushima2008}
K. Fukushima, D. E. Kharzeev, and H. J. Warringa,
Phys. Rev. D \textbf{78}, 074033 (2008).

\bibitem{Tajima2013}
N. Tajima, T. Yamauchi, T. Yamaguchi, M. Suda, 
Y. Kawasugi, H. M. Yamamoto, 
R. Kato, Y. Nishio, and K. Kajita,
Phys. Rev. B \textbf{88}, 075315 (2013).

\bibitem{Hatsugai2015_LL}
Y. Hatsugai, T. Kawarabayashi, and H. Aoki,
Phys. Rev. B \textbf{91}, 085112 (2015).

\bibitem{Hatsugai1993}
Y. Hatsugai and P. A. Lee, Phys. Rev. B \textbf{48}, 4204 (1993).

\bibitem{Nomura2007}
K. Nomura, M. Koshino, and S. Ryu,
Phys. Rev. Lett. \textbf{99}, 146806 (2007).

\bibitem{Kanao2012}
T. Kanao, H. Matsuura, and M. Ogata,
J. Phys. Soc. Jpn. \textbf{81}, 063709 (2012).

\bibitem{Kopnin2008}
N. B. Kopnin, and E. B. Sonin, 
Phys. Rev. Lett. \textbf{100}, 246808 (2008);
Phys. Rev. B \textbf{82}, 014516 (2010).

\bibitem{Mizoguchi2015}
T. Mizoguchi and M. Ogata, 
J. Phys. Soc. Jpn. \textbf{84}, 084704 (2015).

\bibitem{Kozii2017}
V. Kozii and L. Fu, 
arXiv:1708.05841. 

\bibitem{Yoshida2018}
T. Yoshida, R. Peters, and N. Kawakami,
Phys. Rev. B \textbf{98}, 035141 (2018).

\bibitem{Okugawa2019}
R. Okugawa and T. Yokoyama,
Phys. Rev. B \textbf{99}, 041202(R) (2019).

\bibitem{Budich2019}
J. C. Budich, J. Carlstr\"{o}m, F. K. Kunst, and E. J. Bergholtz,
Phys. Rev. B \textbf{99}, 041406(R) (2019).

\bibitem{Yoshida2019}
T. Yoshida, R. Peters, N. Kawakami, and Y. Hatsugai,
Phys. Rev. B \textbf{99}, 121101(R) (2019). 

\bibitem{Papaj2019}
M. Papaj, H. Isobe, and L. Fu,
Phys. Rev. B \textbf{99}, 201107(R) (2019).

\bibitem{Li2017}
D. Li, B. Rosenstein, B. Ya. Shapiro, and I. Shapiro,
Phys. Rev. B \textbf{95}, 094513 (2017).

\bibitem{Hatsugai2011}
Y. Hatsugai and I. Maruyama, 
Europhys. Lett. \textbf{95} 20003 (2011).  

\bibitem{Hatsugai2015}
Y. Hatsugai, K. Shiraishi, and H. Aoki,
New J. Phys. \textbf{17}, 025009 (2015).

\bibitem{Mizoguchi2019}
T. Mizoguchi and Y. Hatsugai,
Europhys. Lett. \textbf{127}, 47001 (2019). 

\bibitem{Mizoguchi2020}
T. Mizoguchi and Y. Hatsugai,
Phys. Rev. B \textbf{101}, 235125 (2020).

\bibitem{Koshino2014}
M. Koshino, T. Morimoto, and M. Sato,
Phys. Rev. B \textbf{90}, 115207 (2014).

\bibitem{Liu2017}
F. Liu, and K. Wakabayashi,
Phys. Rev. Lett. \textbf{118}, 076803 (2017).

\bibitem{Obana2019}
D. Obana, F. Liu, and K. Wakabayashi,
Phys. Rev. B \textbf{100}, 075437 (2019).

\bibitem{Benalcazar2020}
W. A. Benalcazar and A. Cerjan,
Phys. Rev. B \textbf{101}, 161116(R) (2020).

\bibitem{Kudo2019}
K. Kudo, T. Yoshida, and Y. Hatsugai,
Phys. Rev. Lett. \textbf{123}, 196402 (2019). 

\bibitem{remark2}
To be more concrete, for $f_k =0$, $\left(-ig_{k}, 0, |\gamma_k| \right)^{\rm T}$ is a zero mode of $\tilde{h}_k$.


\end{thebibliography}
\end{document}